\documentclass[fp,twocolumn]{jpsj3}
\usepackage{txfonts,color}

\title{Thermoelectric and Magnetic Properties in Doped Fe$_2$VAl Based on a Bipolar Random Anderson Model}

\author{Takami Tohyama$^1$\thanks{tohyama@rs.tus.ac.jp} and Hidetoshi Fukuyama$^2$}
\inst{$^1$Department of Applied Physics, Tokyo University of Science, Katsushika, Tokyo 125-8585, Japan \\
$^2$Research Institute for Science and Technology, Tokyo University of Science, Shinjuku, Tokyo 162-8601, Japan} 

\abst{
We investigate the thermoelectric and magnetic properties of Si-substituted $n$-type and Ti-substituted $p$-type Heusler alloy Fe$_2$VAl using the bipolar random Anderson model, which has been introduced recently to study antisite-defect effects associated with the sign change of the Seebeck coefficient in thermally quenched Fe$_2$VAl. Based on the electronic states of both $n$-type and $p$-type compounds, with the rigid-band shift of the Fermi energy and a temperature-dependent scattering rate taken into account, we elucidate how antisite defects simultaneously influence thermoelectric transport and local magnetic moments. We find that the magnetic moments are enhanced in $n$-type Fe$_2$VAl, whereas they are suppressed in $p$-type Fe$_2$VAl compared with the undoped compound. These contrasting magnetic responses highlight the impact of antisite spin polarization on thermoelectric properties and demonstrate the crucial role of antisite defects in realizing magneto-thermoelectric functionalities in Heusler-type alloys.
}
\begin{document}
\maketitle

\section{Introduction}
\label{Sec1}
The Heusler-type intermetallic Fe$_2$VAl-based compounds are known as promising candidates for thermoelectric materials~\cite{Nishino2011}. Stoichiometric Fe$_2$VAl is nonmagnetic semimetal and the valence-band maximum is located at the $\Gamma$ point, while the conduction-band minimum lies at the X point, with a small band overlap between them~\cite{Guo1998,Singh1998,Weht1998,Weinert1998}, resulting in a pseudogap around the Fermi level $\varepsilon_\mathrm{F}$~\cite{Nishino1997,Okamura2000}. This subtle electronic structure is the origin of the diverse thermoelectric properties induced by element substitution and quenching processes. In Fe$_2$VAl, the resistivity $\rho$ increases as the temperature is lowered and the Seebeck coefficient $S$ exhibits positive values over a wide temperature range~\cite{Nishino2001}, indicating that holes are the majority carriers. Upon Si substitution, in Fe$_2$VAl$_{1-x}$Si$_x$, however, $\rho$ decreases as temperature is lowered and $S$ changes dramatically from positive to negative ($n$-type)  even for small Si concentrations of $x=0.02$ and 0.03~\cite{Kato2001,Lue2007}. This strong sensitivity primarily originates from the introduction of electron carriers, which shifts $\varepsilon_\mathrm{F}$ to higher energies within the pseudogap. On the other hand in the $p$-type Fe$_2$VAl,  Fe$_2$V$_{1-x}$Ti$_x$Al~\cite{Matsuura2002,Slebarski2006,Nakayama2008} realized by substituting Ti for V resulting in holes in the valence band, the resistivity becomes metallic at low temperatures~\cite{Matsuura2002} similar to $n$-type Fe$_2$VAl~\cite{Kato2001,Lue2007}. The $S$ remains positive and its magnitude increases with increasing Ti concentration up to $x=0.03$, and then decreases~\cite{Matsuura2002}. This $p$-type behavior has also been observed in Mn-substituted Fe$_2$V$_{1-x}$Mn$_x$Al~\cite{Jha2024}. As for understanding these transport features in doped Fe$_2$VAl, a phenomenological two-band model has been shown to qualitatively account for the temperature dependence of $S$ in Fe$_2$VAl$_{0.9}$Si$_{0.1}$~\cite{Garmroudi2021}, and evaluations of $S$ in both $n$-type and $p$-type Fe$_2$VAl have also been performed using the linearized Boltzmann transport equation with a constant relaxation time combined with first-principles electronic-structure calculations~\cite{Sato2024}.

Besides the transport properties, magnetic properties have a diversity. In contrast to nonmagnetic Fe$_2$VAl, ferromagnetic order has been observed at low temperatures in Si-substituted $n$-type Fe$_2$VAl~\cite{Jemima2001,Amaladass2015,Tsuji2019}, which has been regarded as a weak itinerant ferromagnet~\cite{Tsuji2019}, since the field-induced magnetic moment is small and weakens with increasing temperature. Furthermore, quenching Fe$_2$VAl$_{0.9}$Si$_{0.1}$ from high temperatures generates a magnetic moment an order of magnitude larger than that of the unquenched sample~\cite{Garmroudi2023b}. Quenched undoped Fe$_2$VAl also induces a magnetic moment~\cite{Garmroudi2022}, but its magnitude is only about one-third of that observed in quenched $n$-type Fe$_2$VAl$_{0.9}$Si$_{0.1}$~\cite{Garmroudi2023b}, highlighting the crucial role of electron doping. Since thermal quenching introduces antisite defects, it is reasonable to attribute such large moments to, for example, V antisites on the Fe sublattice (V$_\mathrm{Fe}$) and Fe antisites on the V sublattice (Fe$_\mathrm{V}$). Hence, magnetic properties are seen to be strongly affected by antisite defects.

The introduction of antisites also causes remarkable changes in $S$: in Fe$_2$VAl, the sign of $S$ becomes negative~\cite{Garmroudi2022}, whereas in $n$-type Fe$_2$VAl$_{0.9}$Si$_{0.1}$, the magnitude of $S$ is reduced~\cite{Garmroudi2023b}. A reduction in $S$ in Fe$_2$VAl$_{1-x}$Si$_x$ has also been reported in thin films~\cite{Hiroi2016} and annealed samples~\cite{Zhang2025}. Regarding $p$-type Fe$_2$VAl, no systematic study has clarified the role of the interplay between antisite defects and their spin polarization. It is therefore necessary to provide a theoretical framework that elucidates the microscopic mechanism of roles played by antisite defects. Hence, systematic understanding based on microscopic analyses in terms of model Hamiltonians is needed to clarify how carrier doping affects the thermoelectric and magnetic properties.

\begin{figure*}[tb]
\center{
\includegraphics[width=0.7\textwidth]{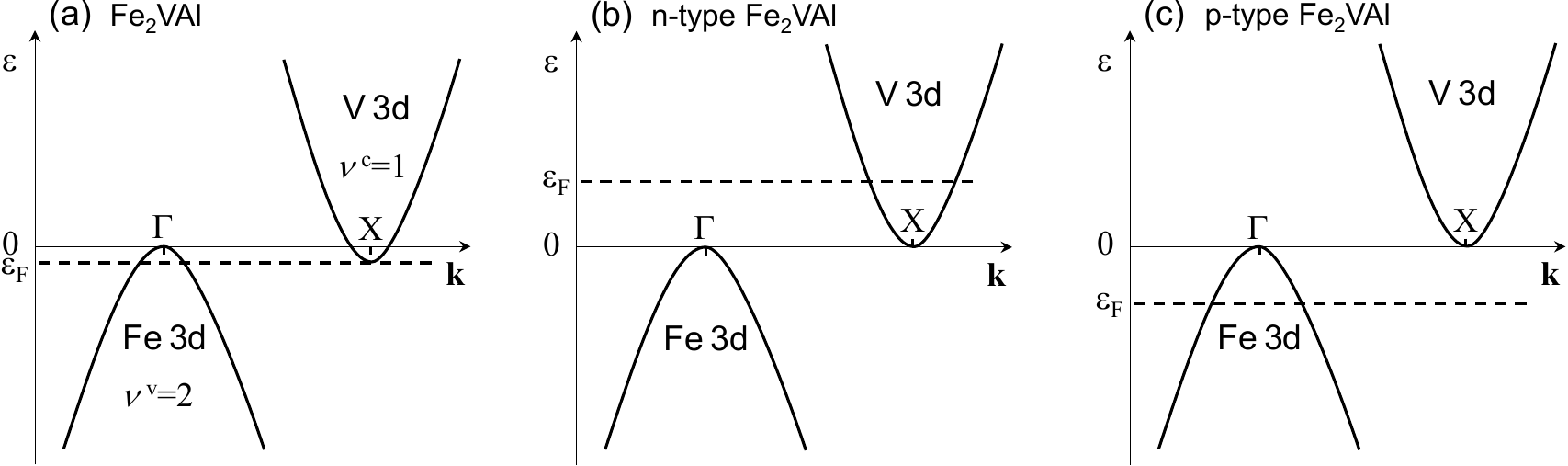}
}
\caption{Schematic illustration of the electronic states in Fe$_2$VAl.
(a) undoped Fe$_2$VAl, (b) $n$-type and (c) $p$-type Fe$_2$VAl. The valence band around the $\Gamma$ point is predominantly composed of Fe $3d$ orbitals with a degeneracy of $\nu^\mathrm{v}=2$, whereas the conduction band near the X point mainly consists of V $3d$ orbitals with a degeneracy of $\nu^\mathrm{c}=1$. The top of the valence band is set as the energy reference. The Fermi energy $\varepsilon_\mathrm{F}$ is indicated by horizontal dashed lines. The energy gap $E_\mathrm{g}$, defined as the energy difference between the valence-band top and the conduction-band bottom, is negative ($E_\mathrm{g}<0$) in (a), whereas $E_\mathrm{g}=0$ in (b) and (c) (see text).}
\label{fig1}
\end{figure*}

In this paper, we theoretically investigate the thermoelectric properties of doped Fe$_2$VAl and the spin polarization associated with antisite defects. We first confirm that a simple rigid-band picture, in which $\varepsilon_\mathrm{F}$ shifts upon carrier doping, captures the overall features of thermoelectric behavior of doped Fe$_2$VAl. To incorporate antisites into our theory, we employ an extended version of the single-impurity Anderson model~\cite{Anderson1961}, the bipolar random Anderson model (BPRAM)~\cite{Tohyama2025}, which has been introduced recently. In this model, V$_\mathrm{Fe}$ and Fe$_\mathrm{V}$ antisites are treated as randomly distributed impurities in the valence and conduction bands, respectively, and induce carrier redistribution between the two bands while conserving the total electron number. This BPRAM framework has successfully explained the sign change of $S$ caused by antisite defects in Fe$_2$VAl~\cite{Tohyama2025}. In $n$-type Fe$_2$VAl, we find that both V$_\mathrm{Fe}$ and Fe$_\mathrm{V}$ antisites are spin-polarized, in contrast to the undoped case where only V$_\mathrm{Fe}$ exhibits spin polarization~\cite{Tohyama2025}. This leads to a magnetic moment approximately twice that of the antisites in undoped Fe$_2$VAl, when the antisite concentration is fixed. The Seebeck coefficient $S$ decreases upon antisite introduction, consistent with experimental observations~\cite{Garmroudi2023b}. In $p$-type Fe$_2$VAl, by contrast, antisite defects lose their magnetic moments because their levels shift above $\varepsilon_{\mathrm{F}}$, resulting in only a small change in $S$. The interplay between thermoelectric and magnetic properties is seen to be governed by the spectral conductivity, which is strongly influenced by spin-polarized antisite defects. These findings provide a new guideline for controlling spin-polarized antisite defects in doped Fe$_2$VAl and demonstrate the crucial role of antisite defects in realizing magneto-thermoelectric functionalities in Heusler-type alloys.

This paper is organized as follows. In Sect.~\ref{Sec2}, we present calculated results for the density of states, spectral conductivity, resistivity, and Seebeck coefficient in doped Fe$_2$VAl. In Sect.~\ref{Sec3}, we introduce the BPRAM and show the electronic states and thermoelectric properties in the presence of antisite defects. Finally, Sect.~\ref{Sec4} discusses the implications of the proposed spin polarization of antisite defects and provides a summary.

\section{Thermoelectric Properties in doped Fe$_2$VAl}
\label{Sec2}
First-principles electronic structure calculations based on the local density approximation or generalized gradient approximation for Fe$_2$VAl have shown a negative band gap, $E_\mathrm{g}<0$, in which the valence-band maximum at $\Gamma$ lies higher in energy than the conduction-band minimum at X~\cite{Guo1998,Singh1998,Weht1998,Weinert1998}. A two-band model with $E_\mathrm{g}=-0.03$~eV and the Fermi energy $\varepsilon_\mathrm{F}$ located at the bottom of the conduction band leads to results consistent with the experimentally observed $S$ and $\rho$~\cite{Tohyama2025}. Figure~\ref{fig1}(a) illustrates this model.

According to our previous work~\cite{Tohyama2025}, we assume an energy dispersion approximated in terms of effective masses. The density of states (DOS) per spin $s$ for the valence (v) and conduction (c) bands is given by
\begin{eqnarray}
D_s^\alpha(\varepsilon)=-\frac{1}{\pi}\mathrm{Im} g_s^\alpha(\varepsilon)
\label{DOS}
\end{eqnarray}
with $\alpha=\mathrm{v}$ or $\mathrm{c}$. The local Green's function $g_s^\alpha(\varepsilon)$ is given by~\cite{Fukuyama1970}
\begin{eqnarray}
g_s^\alpha(\varepsilon)=\left[z_s^\alpha+i\tau_s^\alpha(1+z_s^\alpha)\right]\nu^\alpha/W^\alpha,
\label{LocalG}
\end{eqnarray}
where $W^\alpha$ is the half bandwidth with $W^\mathrm{v(c)}=1.52\ (2.0)$~eV and $\nu^\alpha$ is the band degeneracy with $\nu^\mathrm{v(c)}=2\ (1)$~\cite{Tohyama2025}. Here,
$\tau_s^\alpha=[(1-z_s^\alpha)/(1+z_s^\alpha)]^{1/2}$
and
$z_s^\alpha=\left[\tilde{z}+\Sigma_s^\alpha(\tilde{z})\right]/W^\alpha$, where the self-energy $\Sigma_s^\alpha(\tilde{z})$ arises from hybridization with antisite defects, as described below. By introducing an intrinsic scattering rate $\delta_0$ for a system without antisite defects, the complex energy is defined as
$\tilde{z}=\tilde{\varepsilon}+i\delta_0$
with 
$\tilde{\varepsilon}=\varepsilon+W^\mathrm{v}$
for the valence band and
$\tilde{\varepsilon}=\varepsilon-W^\mathrm{c}-E_\mathrm{g}$
for the conduction band, where the top of the valence band is taken as the origin of energy (see Fig.~\ref{fig1}). 

The spectral conductivity $\sigma_s^\alpha(\varepsilon)$, which governs electrical and heat currents, is given by~\cite{Fukuyama1970}
\begin{eqnarray}
\sigma_s^\alpha(\varepsilon)=(e^2\pi/3)(F^\alpha/W^\alpha)^2L_s^\alpha(\tilde{\varepsilon})
\label{Sigma}
\end{eqnarray}
with
\begin{eqnarray}
L_s^\alpha(\tilde{\varepsilon})=\left(\mathrm{Im} z_s^\alpha\right)^2
-\frac{1}{2}\mathrm{Re}\tau_s^\alpha(1+z_s^\alpha)[\frac{1-(z_s^\alpha)^2}{\mathrm{Im}z_s^\alpha} +3i z_s^\alpha],
\label{L}
\end{eqnarray}
where $e$ is the elementary charge and $F^\alpha$ was determined to reproduces thermoelectric properties in Fe$_2$VAl:  $F^\mathrm{v}=7.0$~eV and $F^\mathrm{c}=11.2$~eV~\cite{Tohyama2025}.

Since the valence-band top and conduction-band bottom are located at different momenta, we assume that there is no direct interplay between the two bands. Under this assumption,the total spectral conductivity is given by $\sigma_s^\mathrm{t}(\varepsilon)=\sigma_s^\mathrm{v}(\tilde{\varepsilon})+\sigma_s^\mathrm{c}(\tilde{\varepsilon})$. In terms of $\sigma_s^\mathrm{t}(\varepsilon)$, the electrical conductivity $L_{11}$ and thermoelectric conductivity $L_{12}$, describing the electrical current density $\mathbf{j}=L_{11}\mathbf{E} + L_{12}(-\nabla T)$ within the linear response theory under an electric field $\mathbf{E}$ and a temperature gradient $\nabla T$~\cite{Aschcroft}, are given by the Sommerfeld-Bethe relationship,
\begin{eqnarray}
L_{11}=\int_{-\infty}^\infty (-\frac{\partial f(\varepsilon)}{\partial \varepsilon}) \sum_s\sigma_s^\mathrm{t}(\varepsilon)d\varepsilon
\label{L11}
\end{eqnarray}
and
\begin{eqnarray}
L_{12}=-\frac{1}{e}\int_{-\infty}^\infty (-\frac{\partial f(\varepsilon)}{\partial \varepsilon})\left(\varepsilon-\mu\right)\sum_s\sigma_s^\mathrm{t}d\varepsilon.
\label{L12}
\end{eqnarray}
Here, $f(\varepsilon)$ is the Fermi distribution function, $f(\varepsilon)=1/(e^{(\varepsilon-\mu)/(k_\mathbf{B}T)}+1)$, where $k_\mathrm{B}$ is the Boltzmann constant. The chemical potential $\mu$ is a temperature-dependent quantity, $\mu=\mu(T)$, determined so as to give the total electron density $n=\sum_s \int_{-\infty}^\infty f(\varepsilon)D_s^\mathrm{t}(\varepsilon)d\varepsilon$ with the total DOS per spin $D_s^\mathrm{t}(\varepsilon)=D_s^\mathrm{v}(\varepsilon)+D_s^\mathrm{c}(\varepsilon)$. The $\rho$ and $S$ are given by $\rho=1/L_{11}$ and $S=L_{12}/L_{11}$, respectively. In general, $L_{12}$ contains contributions from both electrons and phonons~\cite{Ogata2019}; however, the phonon-drag contribution is expected to be small in the present system and is neglected in the following. 

\begin{figure}[tb]
\center{
\includegraphics[width=0.35\textwidth]{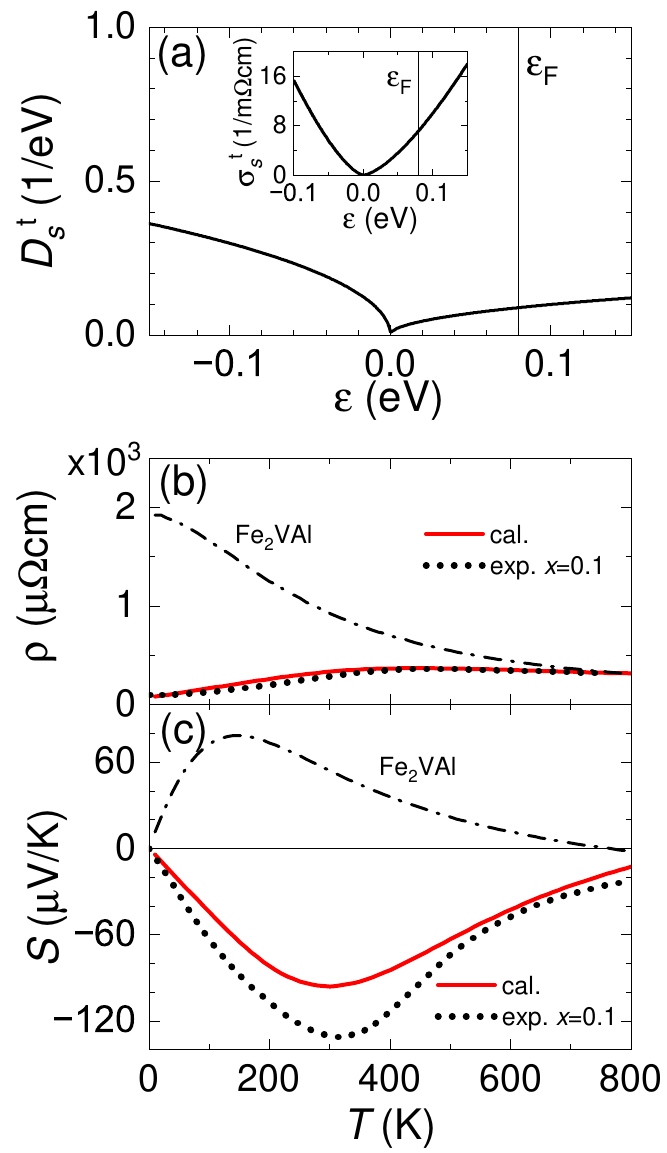}
}
\caption{(Color online) Electronic states and transport properties in $n$-type Fe$_2$VAl.
(a) Total DOS per spin. The energy zero is set at the top of the valence band. The inset shows the zero-temperature total spectral conductivity $\sigma_s^\mathrm{t}$ per spin. The vertical line denotes the position of the Fermi energy $\varepsilon_\mathrm{F}$.
(b) Temperature dependence of the resistivity $\rho$.
(c) Temperature dependence of the Seebeck coefficient $S$.
The red lines represent calculated results. The black dotted lines correspond to experimental data for Fe$_2$VAl$_{1-x}$Si$_x$ with $x=0.1$~\cite{Kato2001}. For comparison, the calculated results for Fe$_2$VAl~\cite{Tohyama2025} are shown by black dash-dotted lines in (b) and (c).}
\label{fig2}
\end{figure}

Si substitution is expected to modify the underlying electronic states. One obvious effect is an increase in electron carriers due to the replacement of Al by Si, which results in a shift of $\varepsilon_\mathrm{F}$ toward higher energies. Another possible effect is a change of $E_\mathrm{g}$. Indeed, an examination of first-principles results for $x=0.1$~\cite{Garmroudi2021} reveals a very small increase in $E_\mathrm{g}$ on the order of 10~meV.  Such an enhancement of $E_\mathrm{g}$ has also been pointed out in a phenomenological two-band analysis of $S$~\cite{Garmroudi2021}. Based on these observations, we take $E_\mathrm{g}=0$ as shown in Fig.~\ref{fig1}(b). Ti-substituted Fe$_2$VAl, on the other hand, introduces hole carriers into the valence band, resulting in a shift of $\varepsilon_\mathrm{F}$ toward lower energies. At the same time, a upward shift of the V-dominated conduction band is expected, since V is replaced by Ti that has a higher 3$d$ energy level. This modification changes $E_\mathrm{g}$, and we take $E_\mathrm{g}=0$, as shown in Fig.~\ref{fig1}(c).

Both $n$-type Fe$_2$VAl$_{1-x}$Si$_x$ and $p$-type Fe$_2$V$_{1-x}$Ti$_x$Al exhibit metallic behavior in $\rho$ at low temperatures for $x \sim 0.1$~\cite{Kato2001,Matsuura2002}. This indicates the presence of a temperature-dependent scattering rate arising from electron-phonon interactions. Accordingly, we incorporate a linear temperature dependence into the scattering rate for the spectral conductivity by replacing $\delta_0$ in $L_s^{\alpha}(\tilde{\varepsilon})$ with $\delta_0 + \gamma_{\mathrm{e\text{-}p}} T$. The parameters $\delta_0$ and $\gamma_{\mathrm{e\text{-}p}}$, together with $\varepsilon_\mathrm{F}$, are determined so as to reproduce the experimentally observed $\rho$ for $x \sim 0.1$~\cite{Kato2001,Matsuura2002}. We obtain $\delta_0 = 0.1$~meV for both $n$-type and $p$-type Fe$_2$VAl. The value of $\gamma_{\mathrm{e\text{-}p}}$ is taken to be $1.38 \times 10^{-6}$~eV/K for the conduction (valence) band and $0.63 \times 10^{-6}$~eV/K for the valence (conduction) band in $n$-type ($p$-type) Fe$_2$VAl, where the band hosting the carriers is assumed to have a larger $\gamma_{\mathrm{e\text{-}p}}$. These values of $\gamma_{\mathrm{e\text{-}p}}$ are comparable to those reported for Ge$_{0.87}$Y$_{0.02}$Sb$_{0.10}$Ag$_{0.01}$Te~\cite{Matsubara2025}. Finally, the effects of magnetic ordering reported experimentally~\cite{Jemima2001,Amaladass2015,Tsuji2019,Slebarski2006} are not considered in the present study, since they are expected to become significant only at low temperatures. In addition, possible contributions from magnetic fluctuations to the thermoelectric properties are beyond the scope of the present work and are left for future investigation.

\begin{figure}[tb]
\center{
\includegraphics[width=0.35\textwidth]{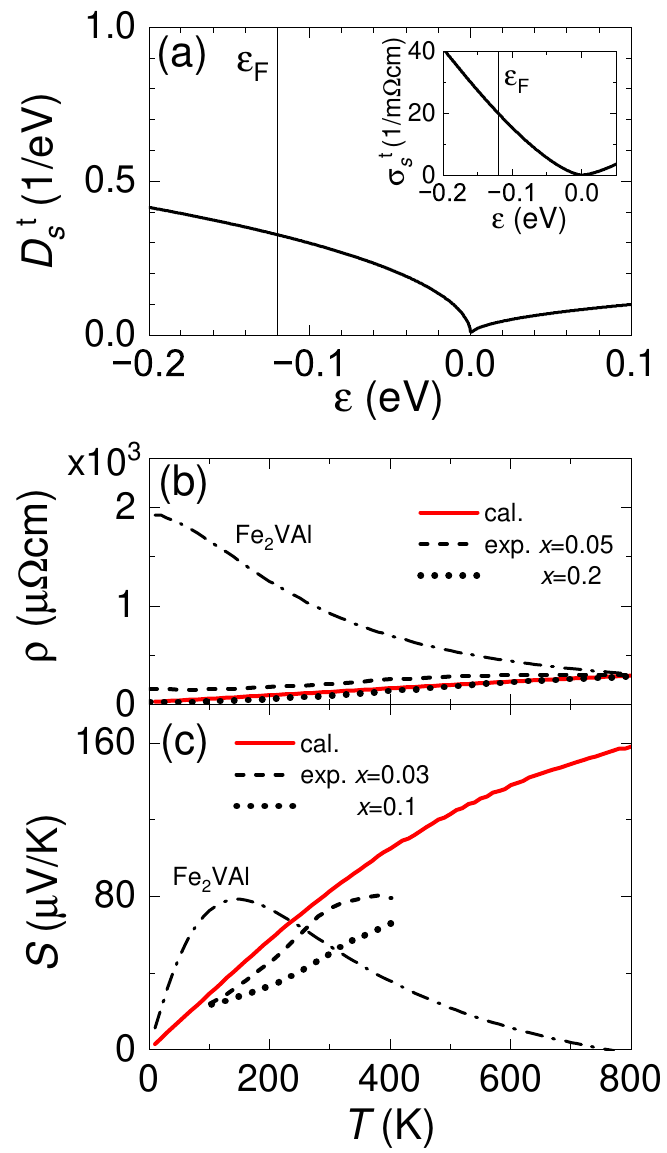}
}
\caption{(Color online) Electronic states and transport properties in $p$-type Fe$_2$VAl.
(a) Total DOS per spin. The energy zero is set at the top of the valence band. The inset shows the zero-temperature spectral conductivity per spin. The vertical line denotes the position of the Fermi energy $\varepsilon_\mathrm{F}$.
(b) Temperature dependence of resistivity $\rho$.
(c) Temperature dependence of Seebeck coefficient $S$. 
The red lines represent calculated results. The black dashed and dotted lines correspond to experimental data for Fe$_2$V$_{1-x}$Ti$_x$Al with $x=0.05$ and $x=0.2$ in (b) and $x=0.03$ and $x=0.1$ in (c)~\cite{Matsuura2002}. For comparison, the calculated results for Fe$_2$VAl~\cite{Tohyama2025} are shown by black dash-dotted lines in (b) and (c).}
\label{fig3}
\end{figure}

Figure~\ref{fig2}(a) shows the total DOS in $n$-type Fe$_2$VAl. The vertical line indicates the position of $\varepsilon_\mathrm{F}$. The zero-temperature spectral conductivity is shown in the inset. Figures~\ref{fig2}(b) and \ref{fig2}(c) present the calculated $\rho$ and $S$, respectively, together with experimental data for $x=0.1$ in Fe$_2$VAl$_{1-x}$Si$_x$~\cite{Kato2001}. The calculated results (red solid lines) qualitatively coincide with the experimental data. The change from positive $S$ in Fe$_2$VAl (black dash-dotted line) to negative values originates from the shift of $\varepsilon_\mathrm{F}$ into the conduction band. When $\varepsilon_\mathrm{F}$ decreases from 0.08~eV, corresponding to a decrease in $x$, the $\rho$ increases and the peak in $S$ shifts to lower temperatures (not shown). This trend is qualitatively consistent with experimental observations~\cite{Kato2001,Lue2007}.

Figure~\ref{fig3}(a) shows the DOS and the zero-temperature spectral conductivity for $p$-type Fe$_2$VAl. Figures~\ref{fig3}(b) and \ref{fig3}(c) present the calculated $\rho$ and $S$, respectively, together with experimental data for Fe$_2$V$_{1-x}$Ti$_x$Al~\cite{Matsuura2002}. It is seen that the calculated $\rho$ (red solid lines) is located between the experimental $x=0.05$ and $x=0.2$ data., while the calculated $S$ is smaller than that in undoped Fe$_2$VAl at low temperatures below $T\sim200$~K and increases with increasing $T$ toward $T=800$~K.  It is of interest to check experimentally the latter trend at high temperatures.

\section{Doped Fe$_2$VAl with antisite defects}
\label{Sec3}
Thermal quenching of Fe$_2$VAl-based compounds induces antisite defects~\cite{Garmroudi2023b,Garmroudi2022,Zhang2025}.  Among these defects, we focus on Fe$_\mathrm{V}$ and V$_\mathrm{Fe}$ antisites, which affect the underlying electronic states in the valence V 3$d$ and conduction Fe 3$d$ band, respectively. An Fe$_\mathrm{V}$ (V$_\mathrm{Fe}$) antisite can be regarded as an impurity in the valence (conduction) band, which is described by the Anderson model~\cite{Anderson1961}. However, since such antisites are randomly distributed within the host sublattice, their effects are more appropriately described by an Anderson-type model with randomly distributed impurities. In addition, charge transfer between an antisite defect and a host band may occur. Assuming that Fe$_\mathrm{V}$ and V$_\mathrm{Fe}$ are present in equal concentrations, charge conservation requires that the number of electrons transferred from Fe$_\mathrm{V}$ antisites to the conduction band equals the number of holes transferred from V$_\mathrm{Fe}$ antisites to the valence band. This charge redistribution can give rise to bipolar effects and substantially reduce the energy gap $E_\mathrm{g}$ making it strongly negative.

To account for this phenomenon, we introduce the BPRAM~\cite{Tohyama2025}, whose Hamiltonian is given by
\begin{eqnarray}
H^\alpha_\mathrm{RAM}&=& \sum_{\mathbf{k},s} \varepsilon^\alpha_\mathbf{k} c^{\alpha\dagger}_{\mathbf{k}s} c^\alpha_{\mathbf{k}s} + \sum_i \left(\varepsilon^\alpha_\mathrm{d} n^\alpha_i + U^\alpha_\mathrm{d} n^\alpha_{i\uparrow}n^\alpha_{i\downarrow}\right) \nonumber \\
 &&+V^\alpha \sum_{\mathbf{k},s,i} \left( d^{\alpha\dagger}_{is} c^\alpha_{\mathbf{k}s} + \mathrm{H.c.}  \right),
\label{HRA}
\end{eqnarray}
where $c^{\alpha\dagger}_{\mathbf{k}s}$ is the creation operator for a mobile electron with momentum $\mathbf{k}$, spin $s$, and dispersion $\varepsilon^\alpha_\mathbf{k}$, $d^{\alpha\dagger}_{is}$ is the creation operator for a localized electron at an antisite $i$, and $n^\alpha_i=n^\alpha_{i\uparrow}+n^\alpha_{i\downarrow}$ with $n^\alpha_{is}=d^{\alpha\dagger}_{is} d^\alpha_{is}$. $\varepsilon^\alpha_\mathrm{d}$ is  the antisite energy level, $U^\alpha_\mathrm{d}$ is on-site Coulomb interaction at the antisites, and $V^\alpha$ represents the hybridization between mobile and antisite electrons. 

Electrons in the valence and conduction bands are scattered by randomly distributed antisite defects via the hybridization term in Eq.~(\ref{HRA}). This scattering induces a $\mathbf{k}$-independent but spin-dependent self-energy for the electrons, denoted by $\Sigma_s^\alpha(\varepsilon)$. By applying a single-site approximation for the antisite defects and employing a self-consistent $T$-matrix approximation for the self-energy, we obtain the following self-consistent equation~\cite{Tohyama2025},
\begin{eqnarray}
\Sigma_s^\alpha(\varepsilon)=c_\mathrm{i}\frac{(V^\alpha)^2G_s^{\mathrm{d},\alpha}(\varepsilon)}{1-(V^\alpha)^2G_s^{\mathrm{d},\alpha}(\varepsilon) g_s^\alpha(\varepsilon)},
\label{SE}
\end{eqnarray}
where $c_\mathrm{i}$ denotes the antisite concentration. Here, $G_s^{\mathrm{d},\alpha}(\varepsilon)$ is the Green's function for antisite states, which is written as
$G_s^{\mathrm{d},\alpha}(\varepsilon)=[\varepsilon-\varepsilon^{\mathrm{d},\alpha}_s-\Sigma_s^{\mathrm{d},\alpha}(\varepsilon)]^{-1}$
with $\Sigma_s^{\mathrm{d},\alpha}(\varepsilon)=(V^\alpha)^2 g_s^\alpha(\varepsilon)$.
The effective antisite energy level is given by
$\varepsilon_s^{\mathrm{d},\alpha}=\varepsilon^\alpha_\mathrm{d}+U^\alpha_\mathrm{d} \left<n^\alpha_{-s}\right>$,
where the spin-dependent antisite occupation number is defined as
$\left<n^\alpha_s\right>=-\frac{1}{\pi}\int_{-\infty}^{\varepsilon_\mathrm{F}}\mathrm{Im} G_s^{\mathrm{d},\alpha}(\varepsilon)d\varepsilon$.

\begin{figure}[tb]
\center{
\includegraphics[width=0.35\textwidth]{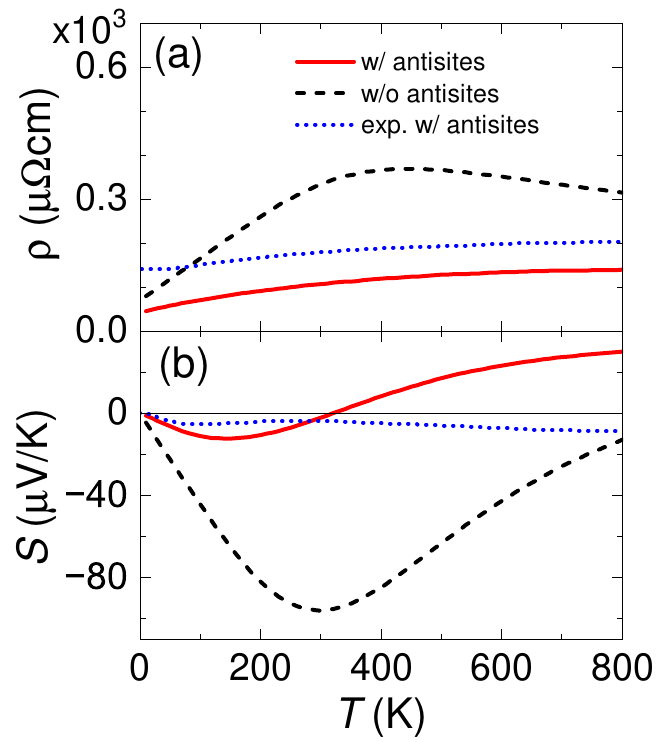}
}
\caption{(Color online) Temperature dependence of (a) the calculated resistivity $\rho$ and (b) the Seebeck coefficient $S$ in the BPRAM for $n$-type Fe$_2$VAl with antisite defects (red solid lines). The blue dashed lines correspond to experimental data for quenched Fe$_2$VAl$_{0.9}$Si$_{0.1}$ containing antisites~\cite{Garmroudi2023b}. For comparison, the results without antisite defects, identical to the red solid lines in Figs.~\ref{fig2}(b) and \ref{fig2}(c), are shown by black dashed lines.}
\label{fig4}
\end{figure}

For the antisite concentration, we assume $c_\mathrm{i}=0.01$, as in our previous work on Fe$_2$VAl~\cite{Tohyama2025}. To satisfy the bipolar condition arising from charge transfer together with carrier doping introduced by Si and Ti substitutions, we set the energy gap and Fermi energy to $E_\mathrm{g}=-0.232$~eV and $\varepsilon_\mathrm{F}=-0.062$~eV for $n$-type Fe$_2$VAl, and $E_\mathrm{g}=-0.28$~eV and $\varepsilon_\mathrm{F}=-0.149$~eV for $p$-type Fe$_2$VAl. For the antisite energy levels, we assume that their energy offsets from the top of the valence band (for $\varepsilon^\mathrm{v}_\mathrm{d}$) and from the bottom of the conduction band (for $\varepsilon^\mathrm{c}_\mathrm{d}$) remain constant and independent of carrier doping. We set this offset to 0.16~eV. This yields $\varepsilon^\mathrm{v}_\mathrm{d}=-0.16$~eV, while $\varepsilon^\mathrm{c}_\mathrm{d}=-0.072$~eV for $n$-type Fe$_2$VAl and $\varepsilon^\mathrm{c}_\mathrm{d}=-0.12$~eV for $p$-type Fe$_2$VAl. Since both $\varepsilon^\mathrm{v}_\mathrm{d}$ and $\varepsilon^\mathrm{c}_\mathrm{d}$ lie below $\varepsilon_\mathrm{F}$ in $n$-type Fe$_2$VAl, we expect spin polarization at both V$_\mathrm{Fe}$ and Fe$_\mathrm{V}$ antisites, using $U_\mathrm{d}^\mathrm{c}=U_\mathrm{d}^\mathrm{v}=0.2$~eV and $V^\mathrm{v}=V^\mathrm{c}=0.1$~eV as in our previous study~\cite{Tohyama2025}. In contrast, no spin polarization is expected for antisite defects in $p$-type Fe$_2$VAl because both $\varepsilon^\mathrm{v}_\mathrm{d}$ and $\varepsilon^\mathrm{c}_\mathrm{d}$ lie above $\varepsilon_\mathrm{F}$. These expectations are indeed confirmed in our calculations, as shown below.
We note that these parameters are chosen as representative values, and our conclusions presented below are robust against small variations in these parameters.

We self-consistently determine the self-energies in Eq.~(\ref{SE}) and calculate $\rho$ and $S$ to examine the effect of antisite defects on the thermoelectric properties of $n$-type and $p$-type Fe$_2$VAl. The red solid lines in Figs.~\ref{fig4}(a) and \ref{fig4}(b) show the calculated $\rho$ and $S$, respectively, for $n$-type Fe$_2$VAl. The resistivity $\rho$ is reduced compared with the value obtained in the absence of antisites and becomes metallic up to $T=800$~K. These changes are consistent with the experimental data for quenched Fe$_2$VAl$_{0.9}$Si$_{0.1}$~\cite{Garmroudi2023b}, shown by the blue dotted line. This reduction in $\rho$ originates from an enhancement of the spectral conductivity due to the increased DOS near $\varepsilon_\mathrm{F}$.

\begin{figure}[tb]
\center{
\includegraphics[width=0.35\textwidth]{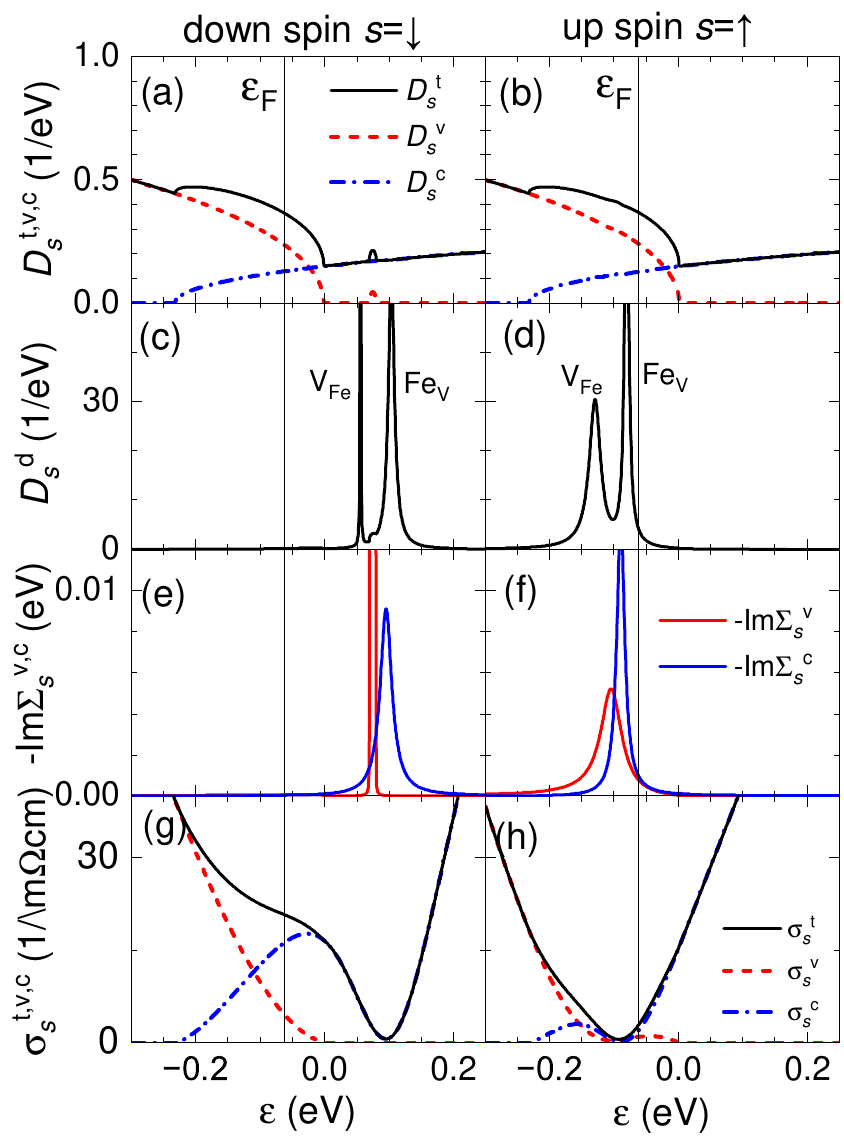}
}
\caption{(Color online) Electronic states of $n$-type Fe$_2$VAl with antisite defects obtained by the BPRAM. The left and right columns correspond to down-spin and up-spin quantities, respectively. (a), (b) The DOS per spin for total electrons (black solid line), valence electrons (red dashed line), and conduction electrons (blue dash-dotted line). (c), (d) The DOS per spin for V$_\mathrm{Fe}$ and Fe$_\mathrm{V}$ antisites. (e), (f) Imaginary part of the spin-dependent self-energy for valence (red solid line) and conduction (blue solid line) electrons. (g), (h) Zero-temperature spectral conductivity per spin for total (black solid line), valence (red dashed line), and conduction (blue dash-dotted line) electrons. The vertical lines indicate the Fermi energy $\varepsilon_\mathrm{F}$.}
\label{fig5}
\end{figure}

To clarify this point, we show the spin-resolved electronic states in the BPRAM in Fig.~\ref{fig5}. Figures~\ref{fig5}(a) and \ref{fig5}(b) display the DOS of the band electrons for each spin. Because of charge transfer, both the valence and conduction bands contribute to the DOS near $\varepsilon_\mathrm{F}$, resulting in an enhancement compared with the case without antisites [see Fig.~\ref{fig2}(a)]. As a result, the total spectral conductivity $\sigma_\uparrow^\mathrm{t} + \sigma_\downarrow^\mathrm{t}$ near $\varepsilon_\mathrm{F}$, shown in Figs.~\ref{fig5}(g) and \ref{fig5}(h), becomes larger than that in the inset of Fig.~\ref{fig2}(a), although the deformation of $\sigma_s^\mathrm{t}$ occurs due to the scattering rate $-\mathrm{Im}\Sigma_s^\alpha$ shown in Figs.~\ref{fig5}(e) and \ref{fig5}(f). This enhanced spectral conductivity leads to the reduction of $\rho$.

The magnitude of $S$ in Fig.~\ref{fig4}(b) is reduced in the presence of antisite defects at low temperatures compared with the case without antisites. This reduction is mainly attributed to the decreased slope of $\sigma_s^\mathrm{t}$ near $\varepsilon_\mathrm{F}$, as shown in Figs.~\ref{fig5}(g) and \ref{fig5}(h), which is caused by antisite scattering. In particular, the conduction-band contribution $\sigma_\downarrow^\mathrm{c}$ around $\varepsilon = 0.15$~eV is suppressed owing to the presence of a peak in $-\mathrm{Im}\Sigma_\downarrow^\mathrm{c}$ at the same energy, leading to a gentler slope of $\sigma_\downarrow^\mathrm{t}$ near $\varepsilon_\mathrm{F}$. The valence-band contribution $\sigma_\uparrow^\mathrm{v}$ around $\varepsilon = -0.1$~eV is also suppressed due to a peak in $-\mathrm{Im}\Sigma_\uparrow^\mathrm{v}$, resulting in a reduced hole contribution to $S$ in the vicinity of $\varepsilon_\mathrm{F}$. These effects contribute to the negative values of $S$ at low temperatures.

With increasing temperature above 300~K, the calculated $S$ in Fig.~\ref{fig4}(b) becomes positive, in contrast to the experimentally observed negative values for quenched Fe$_2$VAl$_{0.9}$Si$_{0.1}$~\cite{Garmroudi2023b}, shown by the blue dotted line. The origin of this discrepancy is unclear, but one possible explanation is a slight modification of $\gamma_\mathrm{e\text{-}p}$ due to the presence of antisite defects in $n$-type Fe$_2$VAl: a reduction of $\gamma_\mathrm{e\text{-}p}$ in the conduction band would enhance the $n$-type carrier contribution at high temperatures.

\begin{figure}[tb]
\center{
\includegraphics[width=0.35\textwidth]{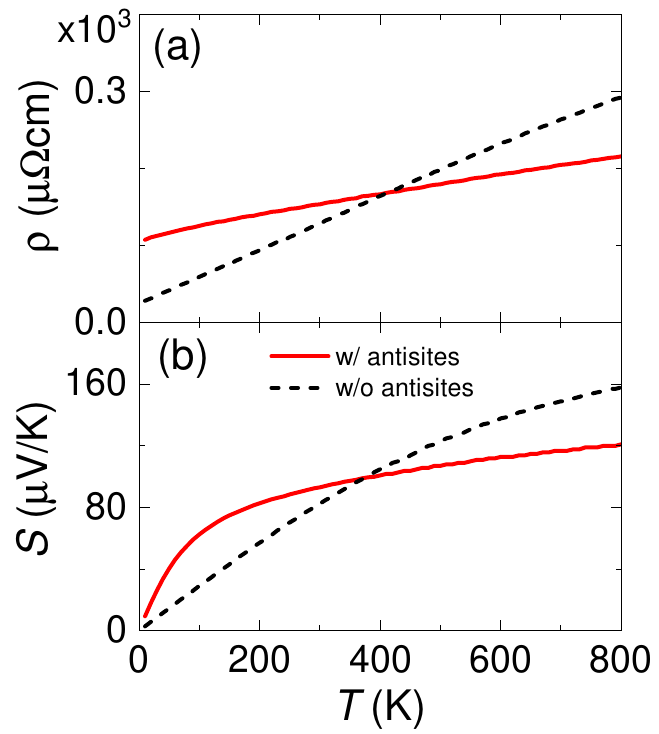}
}
\caption{(Color online) Temperature dependence of (a) the calculated resistivity $\rho$ and (b) the Seebeck coefficient $S$ in the BPRAM for $p$-type Fe$_2$VAl with antisite defects (red solid lines). For comparison, the results without antisite defects, identical to the red solid lines in Figs.~\ref{fig3}(b) and \ref{fig3}(c), are shown by black dashed lines.}
\label{fig6}
\end{figure}

Figures~\ref{fig5}(b) and \ref{fig5}(d) display the DOS of the antisite states, defined as 
$D_s^\mathrm{d} = -\frac{1}{\pi}\mathrm{Im}(G_s^\mathrm{d,v} + G_s^\mathrm{d,c})$.
Both V$_\mathrm{Fe}$ and Fe$_\mathrm{V}$ have spin polarization, since energies of their down- and up-spin states with $\varepsilon_\mathrm{F}$ in between. Consequently, the total magnetic moment, given by the sum of the contributions from V$_\mathrm{Fe}$ and Fe$_\mathrm{V}$, becomes approximately twice as large as that in Fe$_2$VAl, where only V$_\mathrm{Fe}$ antisites are spin-polarized~\cite{Tohyama2025}, when the antisite concentration is the same. This enhancement may account for the experimental observation that Si-substituted samples exhibit larger antisite-induced magnetic moments~\cite{Garmroudi2023b}.

\begin{figure}[tb]
\center{
\includegraphics[width=0.35\textwidth]{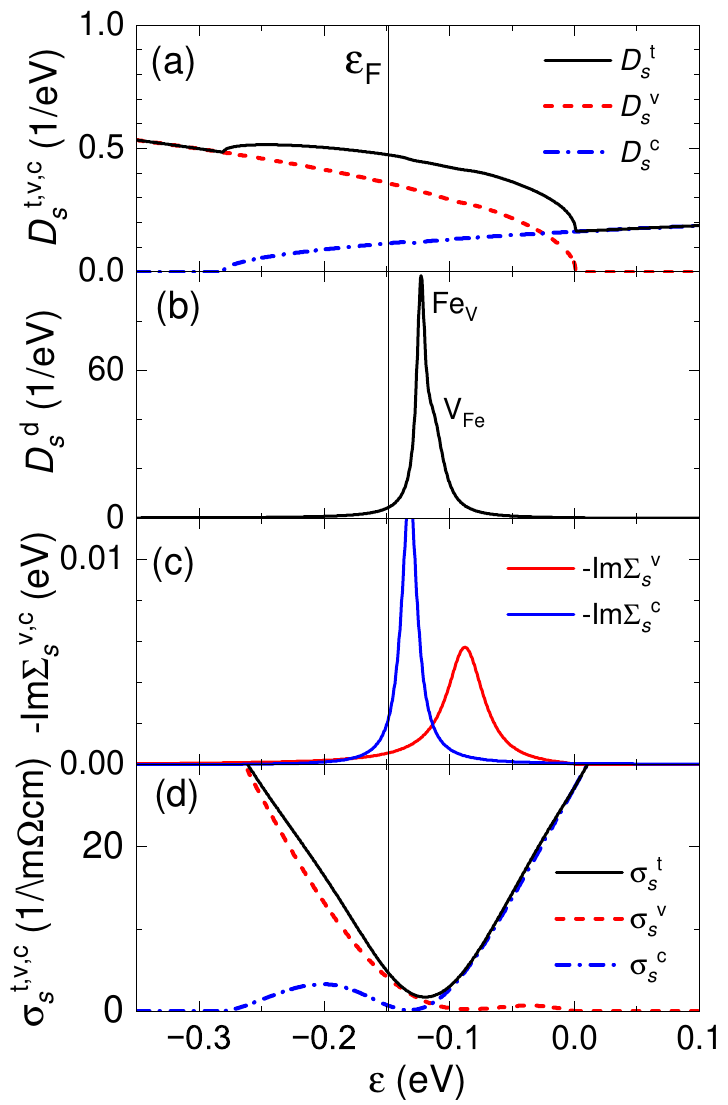}
}
\caption{(Color online) Electronic states of $p$-type Fe$_2$VAl with antisite defects obtained by the BPRAM. No spin polarization is present. 
(a) DOS per spin for the total electrons (black solid line), valence electrons (red dashed line), and conduction electrons (blue dash-dotted line). 
(b) DOS per spin for V$_\mathrm{Fe}$, which produces a shoulder, and for Fe$_\mathrm{V}$, which corresponds to a peak. 
(c) Imaginary part of the self-energy for the valence (red solid line) and conduction (blue solid line) electrons. 
(d) Zero-temperature spectral conductivity per spin for the total (black solid line), valence (red dashed line), and conduction (blue dash-dotted line) electrons. 
The vertical line represents the Fermi energy $\varepsilon_\mathrm{F}$.
}
\label{fig7}
\end{figure}

\begin{figure}[tb]
\center{
\includegraphics[width=0.45\textwidth]{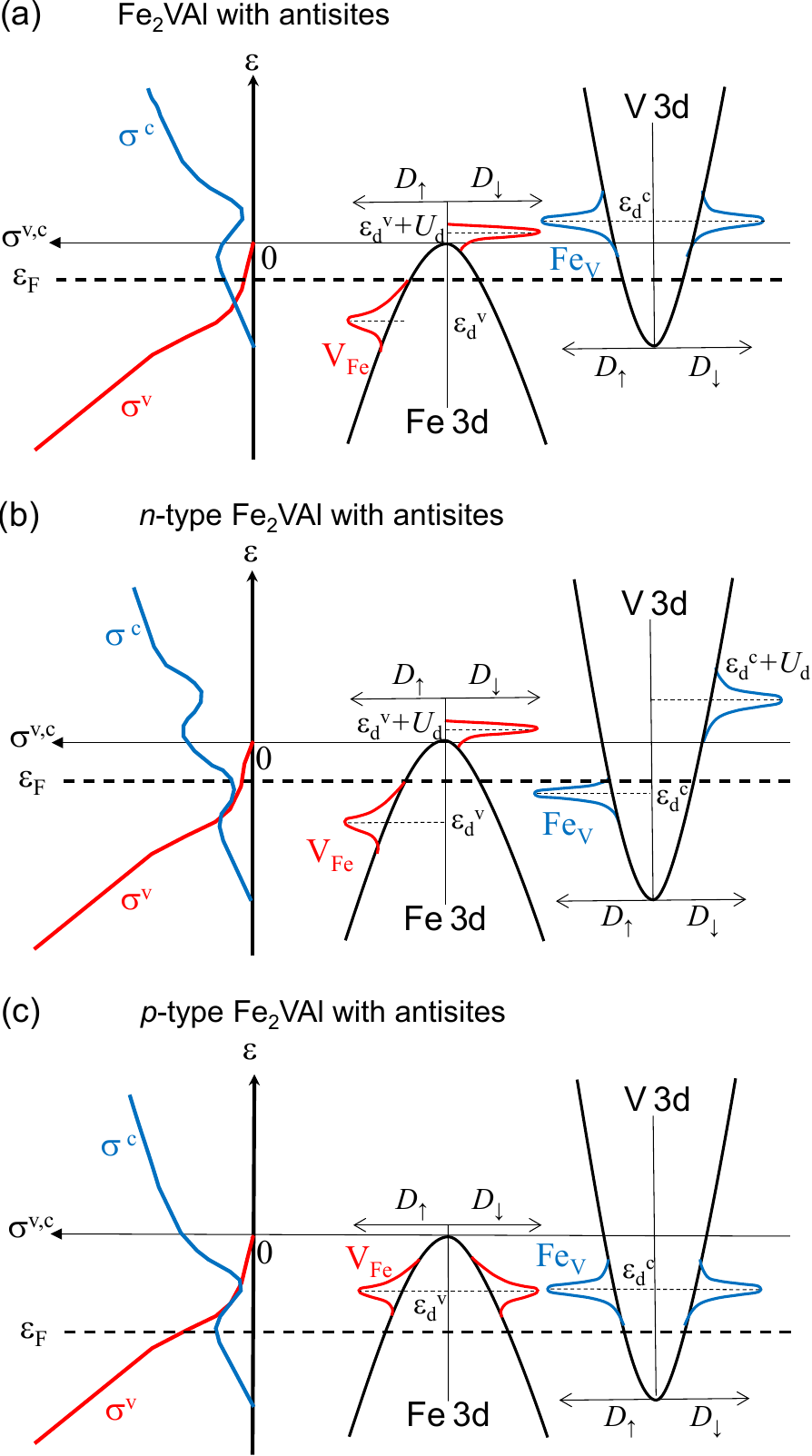}
}
\caption{(Color online) Schematic illustration of the electronic states and the spectral conductivity in Fe$_2$VAl with antisite defects. 
(a) Undoped Fe$_2$VAl, (b) $n$-type, and (c) $p$-type Fe$_2$VAl. 
The top of the valence band is set as the energy reference. 
The Fermi energy $\varepsilon_\mathrm{F}$ is indicated by horizontal dashed lines. 
The left side of the vertical energy axis represents the spectral conductivity averaged over spins for the valence band, $\sigma^\mathrm{v}(\varepsilon)$ (red solid curve), and for the conduction band, $\sigma^\mathrm{c}(\varepsilon)$ (blue solid curve).  The electronic states are shown on the right side of the vertical energy axis. The valence band is composed of Fe $3d$ orbitals, whereas the conduction band consists of V $3d$ orbitals. 
The DOS per spin ($D_\uparrow$ and $D_\downarrow$) for the valence and conduction bands is shown by black solid lines. 
The antisite DOS for V$_\mathrm{Fe}$ and Fe$_\mathrm{V}$ is depicted by red and blue solid lines, respectively.
In (a), the energy level of V$_\mathrm{Fe}$, denoted by $\varepsilon_\mathrm{d}^\mathrm{v}$, appears below $\varepsilon_\mathrm{F}$ for the $\uparrow$-spin antisite state, while the $\downarrow$-spin state is split upward by the on-site Coulomb interaction $U_\mathrm{d}$. 
Both spin states of Fe$_\mathrm{V}$ appear above $\varepsilon_\mathrm{F}$ at $\varepsilon_\mathrm{d}^\mathrm{c}$. 
In (b), the spin states of Fe$_\mathrm{V}$ are also split by $U_\mathrm{d}$, whereas in (c) no splitting occurs for either V$_\mathrm{Fe}$ or Fe$_\mathrm{V}$.}
\label{fig8}
\end{figure}

Figure~\ref{fig6} shows the calculated $\rho$ and $S$ for $p$-type Fe$_2$VAl with antisite defects. The magnitude of $\rho$ does not change significantly, although its slope is modified and its value at $T < 400$~K increases compared with the case without antisites. The magnitude of $S$ at $T < 400$~K is also enhanced in the presence of antisites. To confirm these predictions, systematic experiments that control the antisite concentration in $p$-type Fe$_2$V$_{1-x}$Ti$_x$Al are desired in the future.

The corresponding electronic states are shown in Fig.~\ref{fig7}. The Fermi energy $\varepsilon_\mathrm{F}$ lies inside the valence band, as shown in Fig.~\ref{fig7}(a). No spin polarization appears, since the DOS of V$_\mathrm{Fe}$ and Fe$_\mathrm{V}$ is located above $\varepsilon_\mathrm{F}$ [see Fig.~\ref{fig7}(b)]. In Fig.~\ref{fig7}(d), we find that $\sigma_s^\mathrm{c}$ near $\varepsilon_\mathrm{F}$ is strongly suppressed due to the large $-\mathrm{Im}\Sigma_s^\mathrm{c}$ shown in Fig.~\ref{fig7}(c). This suppression is the origin of the enhancement of $\rho$ and $S$ at low temperatures in Fig.~\ref{fig6}.

\section{Summary and discussions}
\label{Sec4}

Spin polarization of antisite defects gives rise to magnetic moments in Fe$_2$VAl. In our previous work~\cite{Tohyama2025}, we found that V$_\mathrm{Fe}$ is spin-polarized whereas Fe$_\mathrm{V}$ is not in undoped Fe$_2$VAl. In this paper, we show that both V$_\mathrm{Fe}$ and Fe$_\mathrm{V}$ are spin-polarized in $n$-type Fe$_2$VAl, while neither of them is polarized in $p$-type Fe$_2$VAl, as discussed in Sec.~\ref{Sec3}. This difference originates from the relative positions of the antisite energy levels with respect to $\varepsilon_\mathrm{F}$, which depend on whether the system is $n$-type or $p$-type. 

Figure~\ref{fig8} summarizes the schematic electronic states in Fe$_2$VAl with antisite defects (the right side of the vertical energy axis). From this illustration, we can predict that antisite-induced magnetic moments decrease from $n$-type to $p$-type Fe$_2$VAl, provided that the antisite concentration is the same. The recent experimental observation that Si-substituted samples exhibit larger antisite-induced magnetic moments compared with the undoped case~\cite{Garmroudi2023b} may reflect this behavior. It is also interesting to note that the magnetic moments in $p$-type Fe$_2$V$_{1-x}$Mn$_x$Al decrease with increasing $x$~\cite{Jha2024}. If the magnetic moment at $x=0$ originates from antisite defects, a possible source of the reduction may be the shift of $\varepsilon_\mathrm{F}$ deeper into the valence band, as illustrated in Fig.~\ref{fig8}(c). Further systematic investigations of the effects of antisite defects in doped Fe$_2$VAl are desired in the future.

In the left side of the vertical energy axis in Fig.~\ref{fig8}, the spectral conductivity for the valence (conduction) band, $\sigma^{\mathrm{v(c)}}(\varepsilon) = \sigma^{\mathrm{v(c)}}_{\uparrow}(\varepsilon) + \sigma^{\mathrm{v(c)}}_{\downarrow}(\varepsilon)$, is shown by red (blue) solid curves. These results highlight the interplay between thermoelectric and magnetic properties in Fe$_2$VAl with antisite defects. In Fig.~\ref{fig8}(a), $\sigma^{\mathrm{v}}(\varepsilon)$ near $\varepsilon_{\mathrm{F}}$ is suppressed and flattened due to the spin-$\uparrow$ level of the spin-polarized V$_{\mathrm{Fe}}$ antisite. These changes reduce the hole contribution to $S$, resulting in a negative Seebeck coefficient, consistent with the behavior reported for undoped Fe$_2$VAl~\cite{Tohyama2025}. In $n$-type Fe$_2$VAl, shown in Fig.~\ref{fig8}(b), the spin-$\uparrow$ level of spin-polarized Fe$_{\mathrm{V}}$ suppresses and flattens $\sigma^{\mathrm{c}}(\varepsilon)$. Consequently, $\sigma^{\mathrm{v}}(\varepsilon)$ and $\sigma^{\mathrm{c}}(\varepsilon)$ take similar values near $\varepsilon_{\mathrm{F}}$, leading to a small Seebeck coefficient at low temperatures, as demonstrated in Fig.~\ref{fig4}(b). In $p$-type Fe$_2$VAl, all antisite levels lie above $\varepsilon_{\mathrm{F}}$, and their influence on the spectral conductivity near $\varepsilon_{\mathrm{F}}$ is weak, as shown in Fig.~\ref{fig6}(b). These close relationships between spectral conductivity and spin-polarized electronic states demonstrate the crucial role of antisite defects in controlling both thermoelectric and magnetic properties of Heusler-type alloys. Dynamical effects associated with fluctuations of antisite magnetic moments, which are not taken into account in the present study, may further contribute to the modification of $S$.

\begin{figure}[tb]
\center{
\includegraphics[width=0.35\textwidth]{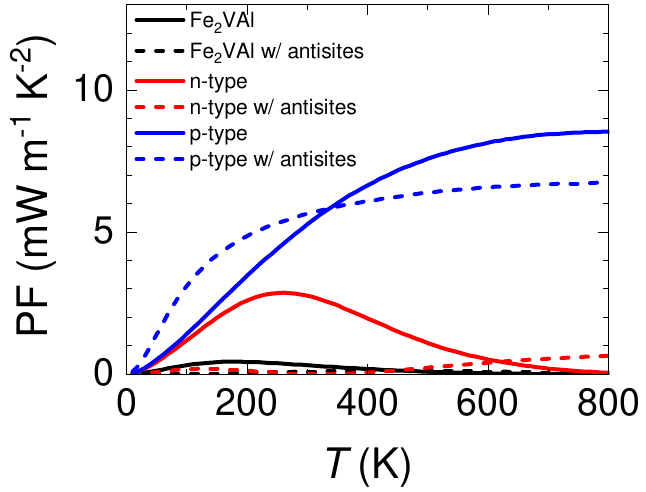}
}
\caption{(Color online) Calculated power factor (PF) of undoped and doped Fe$_2$VAl with (solid lines) and without (dashed lines) antisite defects. The black, red, and blue lines represent the results for undoped, $n$-type, and $p$-type Fe$_2$VAl, respectively.}
\label{fig9}
\end{figure}

The power factor (PF), defined as $S(T)^2/\rho(T)$, serves as a guideline for evaluating the energy conversion efficiency of thermoelectric materials. Figure~\ref{fig9} summarizes the calculated PF for undoped and doped Fe$_2$VAl with and without antisite defects. In the absence of antisites, PF increases with carrier doping. This enhancement is attributed to both the decrease in $\rho$ and the increase in the absolute value of $S$, as shown in Figs.~\ref{fig2} and \ref{fig3}. The PF of $p$-type Fe$_2$VAl is larger than that of $n$-type Fe$_2$VAl. This trend is opposite to experimental data~\cite{Matsuura2002}. This difference arises because the calculated $S$ is larger and smaller than the experimental values for $p$-type and $n$-type Fe$_2$VAl, respectively. 

With antisites, the calculated PF of undoped Fe$_2$VAl~\cite{Tohyama2025} decreases, in contrast to the experimentally observed large enhancement of PF in fully quenched Fe$_2$VAl with large antisite concentration~\cite{Garmroudi2022}. This discrepancy primarily arises from the small magnitude of the calculated $S$, which reflects the low antisite concentration assumed in our model~\cite{Tohyama2025}. In addition, dynamical contributions from fluctuations of antisite magnetic moments may lead to a further enhancement of $S$. In $n$-type Fe$_2$VAl, the introduction of antisite defects reduces the PF, which is consistent with experimental behavior~\cite{Garmroudi2023b}. In $p$-type Fe$_2$VAl, the PF exhibits nearly the same magnitude with and without antisites. This small change is attributed to the absence of spin-polarized antisites, which leads to weaker scattering of hole carriers near $\varepsilon_\mathrm{F}$.

In summary, we have investigated the thermoelectric properties and spin polarization associated with antisite defects in doped Fe$_2$VAl. We have demonstrated that, in doped Fe$_2$VAl, both the rigid-band shift of the Fermi energy $\varepsilon_\mathrm{F}$ and the temperature-dependent scattering rate play crucial roles in understanding the temperature dependence of the Seebeck coefficient and the resistivity based on the Sommerfeld-Bethe relationship. In the presence of V$_\mathrm{Fe}$ and Fe$_\mathrm{V}$ antisite defects, the BPRAM with a shifted $\varepsilon_\mathrm{F}$ yields spin polarization at both types of antisites in $n$-type Fe$_2$VAl, giving rise to a magnetic moment that is approximately twice as large as that of antisites in undoped Fe$_2$VAl,  when the antisite concentration is the same. At the same time, the Seebeck coefficient is reduced in the presence of antisites. These results are consistent with existing experimental observations. In contrast, in $p$-type Fe$_2$VAl, spin polarization of antisites diminishes because the Fermi energy lies deep inside the valence band and the antisite levels are located above $\varepsilon_\mathrm{F}$. These theoretical results demonstrate the impact of carrier substitution on the spin polarization of antisite defects in Fe$_2$VAl and indicate the crucial role of antisites in realizing magneto-thermoelectric functionalities in Heusler-type alloys. Constructing a theoretical framework that takes into account the dynamical effects of antisite magnetic moments remains a challenge for future studies aiming to understand magneto-thermoelectric properties.

\end{document}